%% file: MainarXiv.tex
\documentclass[%
reprint,
superscriptaddress,
amsmath,amssymb,
aps,prl,
twocolumn,
showpacs,
]{revtex4-1}
\usepackage{tikz}
\usepackage{graphicx}
\usepackage{dcolumn}
\usepackage{bm}
\usepackage{braket}
\usepackage{hyperref}

\usepackage{amsmath}
\usepackage{graphicx}
\usepackage{fancyhdr}
\usepackage{float}
\usepackage{todonotes}
\usepackage{siunitx}
\usepackage{xcolor}
\usepackage{xspace}
\usepackage{eucal}
\usepackage{todonotes}
\usepackage{multirow}
\usepackage[utf8]{inputenc}
\usepackage{pgfplots}
\pgfplotsset{compat=1.14}
\usepackage{upgreek}
\usepackage{siunitx}
\usepackage{gensymb}
\usepackage{amsmath}
\usepackage{caption}
\usepackage{subcaption}
\usepackage{textcomp}
\usepackage{comment}
\usepackage{siunitx}
\usepackage[justification=justified,singlelinecheck=false]{caption}
\usepackage{todonotes}
\usepackage{cleveref}

\Crefname{figure}{Figure}{Figures}


\begin{document}
\title{A compact high-precision periodic-error-free heterodyne interferometer}
%
\author{Ki-Nam Joo}
\affiliation{College of Optical Sciences, University of Arizona, 1630 E. University Blvd., Tucson, AZ 85721, USA}
\affiliation{Dept. of Photonic Engineering, Chosun University, 309 Pilmun-daero, Dong-gu, Gwangju, 61452, South Korea}
\author{Erin Clark}
\author{Yanqi Zhang}
\affiliation{College of Optical Sciences, University of Arizona, 1630 E. University Blvd., Tucson, AZ 85721, USA}
\author{Jonathan D. Ellis}
\affiliation{College of Optical Sciences, University of Arizona, 1630 E. University Blvd., Tucson, AZ 85721, USA}
\affiliation{Clerio Vision Inc., 1892 S. Winton Rd., Rochester, NY 14618, USA}
\author{Felipe~Guzman}\email[Electronic mail: ]{felipe@optics.arizona.edu}
\affiliation{College of Optical Sciences, University of Arizona, 1630 E. University Blvd., Tucson, AZ 85721, USA}%
\date{\today}
%
\begin{abstract}
We present the design, bench-top setup, and experimental results of a compact heterodyne interferometer that achieves picometer-level displacement sensitivities in air over frequencies above 100~mHz. The optical configuration with spatially separated beams prevents frequency and polarization mixing, and therefore eliminates periodic errors. The interferometer is designed to maximize common-mode optical laser beam paths to obtain high rejection of environmental disturbances, such as temperature fluctuations and acoustics. The results of our experiments demonstrate the short- and long-term stabilities of the system during stationary and dynamic measurements. In addition, we provide measurements that compare our interferometer prototype with a commercial system, verifying our higher sensitivity of 3\,pm, higher thermal stability by a factor of two, and periodic-error-free performance.  
\end{abstract}
\maketitle
\section{Introduction}

Displacement measuring interferometry (DMI) is a crucial technology to observe dynamic systems and, for example, determine the precise position of a moving stage, measure position errors of a target, and control the motion of a precise machine \cite{1Steinmetz}. DMI has the advantages of non-contact, traceable, accurate measurements with high dynamic range \cite{2Bobroff}. Although environmental factors such as refractive index change of air impact the measurement performance, DMI has become indispensable for dimensional metrology because of the capability of reaching sub-nanometer precision \cite{3Pierce,4Guzman}. DMI systems operated in a vacuum environment are likely intended for long-term usage, which means that its long-term stability is essential. In these cases, the interferometer should be designed in a way that the impact of noise effects that are prominent over large time scales (low frequencies), such as temperature, is low to prevent long-term drifts in the measurement results. To this end, it is possible to design interferometer topologies that allow differential measurements and provide common-mode optical path lengths that provide high rejection ratios to such disturbances \cite{5Lu,6Soomaergren}.

Furthermore, heterodyne laser interferometry is less sensitive to source intensity fluctuations because its measurement techniques are based on phase-locked-loops, lock-in detection or discrete Fourier transforms \cite{7Wand,8Heinzel}, and provides a large dynamic range as well as unambiguous directionality. However, its inherent use of two laser frequencies leads to systematic periodic errors, which are typically at a level of 1-10~nm \cite{9Bobroff,10Hou,11Xie}. Significant effort has been dedicated to analyze and compensate for such periodic errors \cite{12Wu,13Eom,14Deng,15Ellis,16Fu,17Wang}, for example by feedback control schemes and system modifications. Spatial separation of the two beams with different frequencies has been proposed to fundamentally eliminate the polarization and frequency mixing, which results in sub-nm accuracy. However, the design of heterodyne laser interferometers without periodic errors that would allow differential measurements from common-mode paths, is difficult because of the necessary spatial separation between the beams. Typically, periodic-error-free heterodyne interferometers that achieve high displacement sensitivity utilize ultra-stable optomechanical baseplates made of materials such as Zerodur \textsuperscript{\textregistered} and ULE \cite{4Guzman,18Schuldt}, to affix the interferometer optical components.

In this investigation, we present a dimensionally stable compact heterodyne laser interferometer that includes spatially separated beams, as well as common-mode differential optical paths. A so-called \emph{reference} interferometer measures the system phase noise while a \emph{measurement} interferometer measures the displacement of the target. Thus, it is possible to reduce the coupling of disturbances that commonly affect the two interferometers, such as thermal drift, and acoustics, as well as their impact on displacement measurements.

\section{Principle of operation}
\begin{figure}[htbp]
\centering
\fbox{\includegraphics[width=.95\linewidth]{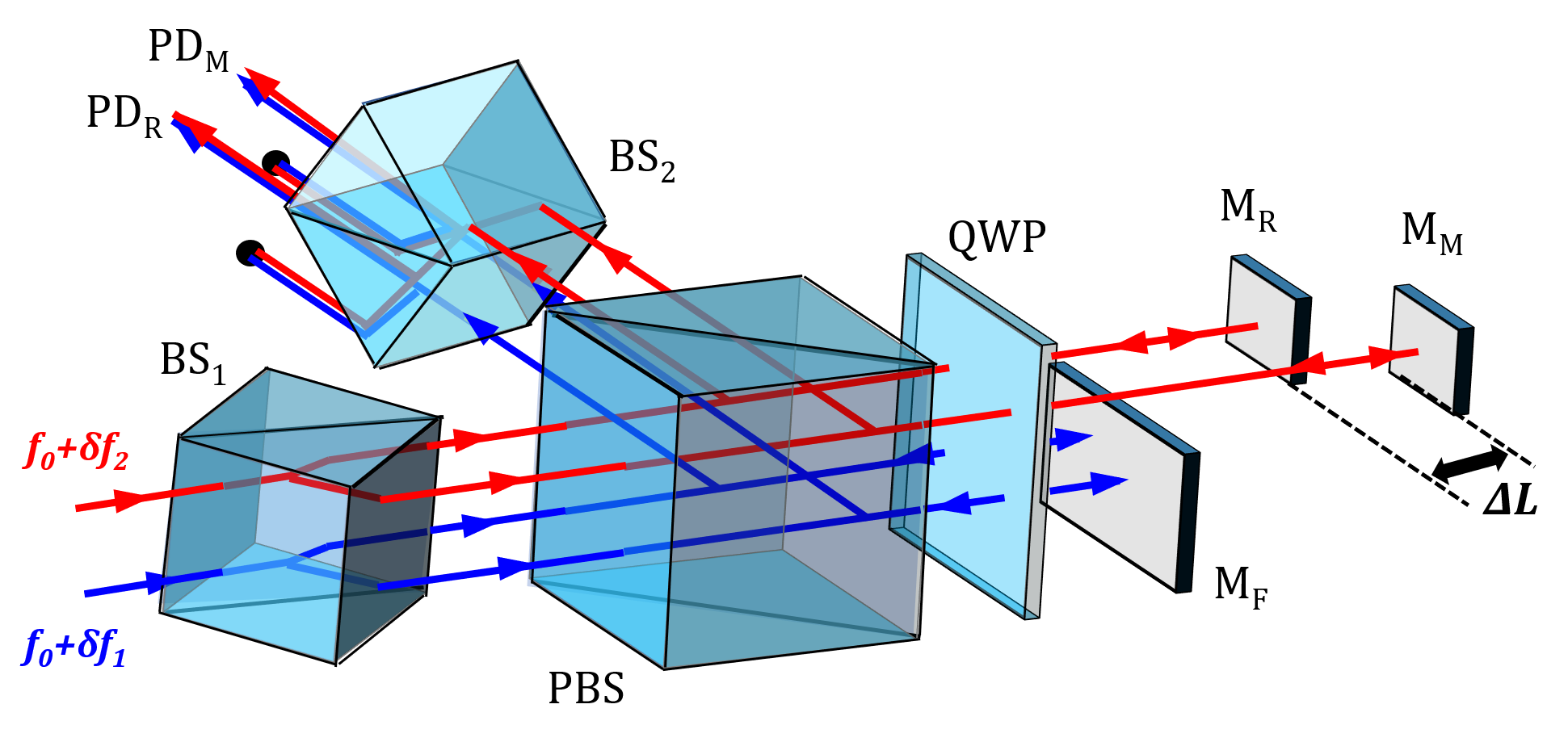}}
\caption{Optical configuration of the proposed differential heterodyne laser interferometer with spatial beam separation; $\textnormal{BS}_1$, 45{\degree} tilted non-polarizing beam splitter; PBS, polarizing beam splitter; $\textnormal{BS}_2$, 45{\degree} tilted and axially 90{\degree} rotated non-polarizing beam splitter; QWP, 45{\degree} rotated quarter wave plate; $\textnormal{M}_\textnormal{F}$, fixed mirror, $\textnormal{M}_\textnormal{R}$, reference mirror; $\textnormal{M}_\textnormal{M}$, measurement mirror; $\textnormal{PD}_\textnormal{R}$, reference photodetector; $\textnormal{PD}_\textnormal{M}$, measurement photodetector.}
\label{Fig1}
\end{figure}

\Cref{Fig1} shows the optical layout of our heterodyne laser interferometer. We use two acousto-optic frequency shifters (AOFSs), (not shown in \Cref{Fig1}), to generate two spatially separated beams from a common source, at slightly different frequencies, ($f_0+\delta f_1$,$f_0+\delta f_2$), where $f_0$ is the optical frequency of the main laser source, $\delta f_1$ and $\delta f_2$ are the corresponding RF frequency shifts. These two beams are fiber-coupled and delivered to the interferometer with a separation of 15~mm along the vertical axis. The beams pass through a 45{\degree} tilted non-polarizing beam splitter ($\textnormal{BS}_\textnormal{1}$), and are divided into horizontally separated beams with a separation of 15~mm, as shown in \Cref{Fig1}. This yields four parallel beams that are geometrically shifted vertically and horizontally in a square-like array, that pass through a polarizing beam splitter (PBS) and a 45{\degree} rotated quarter-waveplate (QWP) toward a fixed mirror ($\textnormal{M}_\textnormal{F}$), a reference mirror ($\textnormal{M}_\textnormal{R}$), and a measurement mirror ($\textnormal{M}_\textnormal{M}$), respectively. The two horizontally separated beams at a frequency of ($f_0+\delta f_1$) are reflected from $\textnormal{M}_\textnormal{F}$ and the two beams at frequency ($f_0+\delta f_2$) are reflected from $\textnormal{M}_\textnormal{R}$ and $\textnormal{M}_\textnormal{M}$, respectively. The four beams travel back along their foregoing paths into the PBS, passing again through the QWP and are then reflected by the PBS, from their change in polarization, to go toward the non-polarizing recombination beam splitter ($\textnormal{BS}_\textnormal{2}$), which is axially 90{\degree} rotated and tilted by 45{\degree}. The orientation of $\textnormal{BS}_\textnormal{2}$ allows the spatially separated laser interferometer beams to recombine and generate the corresponding beat notes measured at the photodetectors, $\textnormal{PD}_\textnormal{R}$ and $\textnormal{PD}_\textnormal{M}$. These beat notes are the reference and measurement signals according to their corresponding reflections at $\textnormal{M}_\textnormal{R}$ and $\textnormal{M}_\textnormal{M}$, respectively. By measuring the phase difference between them, we can thus observe the relative displacement ($\Delta L$) between the reference mirror, $\textnormal{M}_\textnormal{R}$, and the measurement target, $\textnormal{M}_\textnormal{M}$.

\subsection{Thermal Stability}
It is possible to design the optical path lengths of a laser interferometer to conduct differential measurements, thus increasing its measurement stability. In this case, the laser-interferometric phase change driven by the temperature-induced dimensional changes of the optical components becomes common-mode and is mitigated when the optical paths of the interfering beams are similar enough, with the exception of the non-common segment that measures the target dynamics. Interferometer topologies that allow common optical phases can be easily designed when the two laser beams are coaxial as in commercial double-pass plane mirror interferometers. However, achieving this becomes more challenging with spatially separated beams because the optical axes are intentionally split and traversing different paths to eliminate the periodic errors. Several types of heterodyne laser interferometers with spatially separated beams \cite{19Wu,20Joo,21Joo,22Hu} have been reported, which targeted reductions of periodic errors, but not necessarily high stability. However, temperature fluctuations become a dominant factor once environmental effects and periodic errors are minimized. Therefore, we designed our interferometer to minimize the phase noise caused by temperature fluctuations, although the optical paths are not perfectly common. It is possible to account for the effect of temperature fluctuations on the interferometric phase measurement by building geometrical relationships between the four beams.

\begin{figure}[htbp]
\centering
\fbox{\includegraphics[width=.95\linewidth]{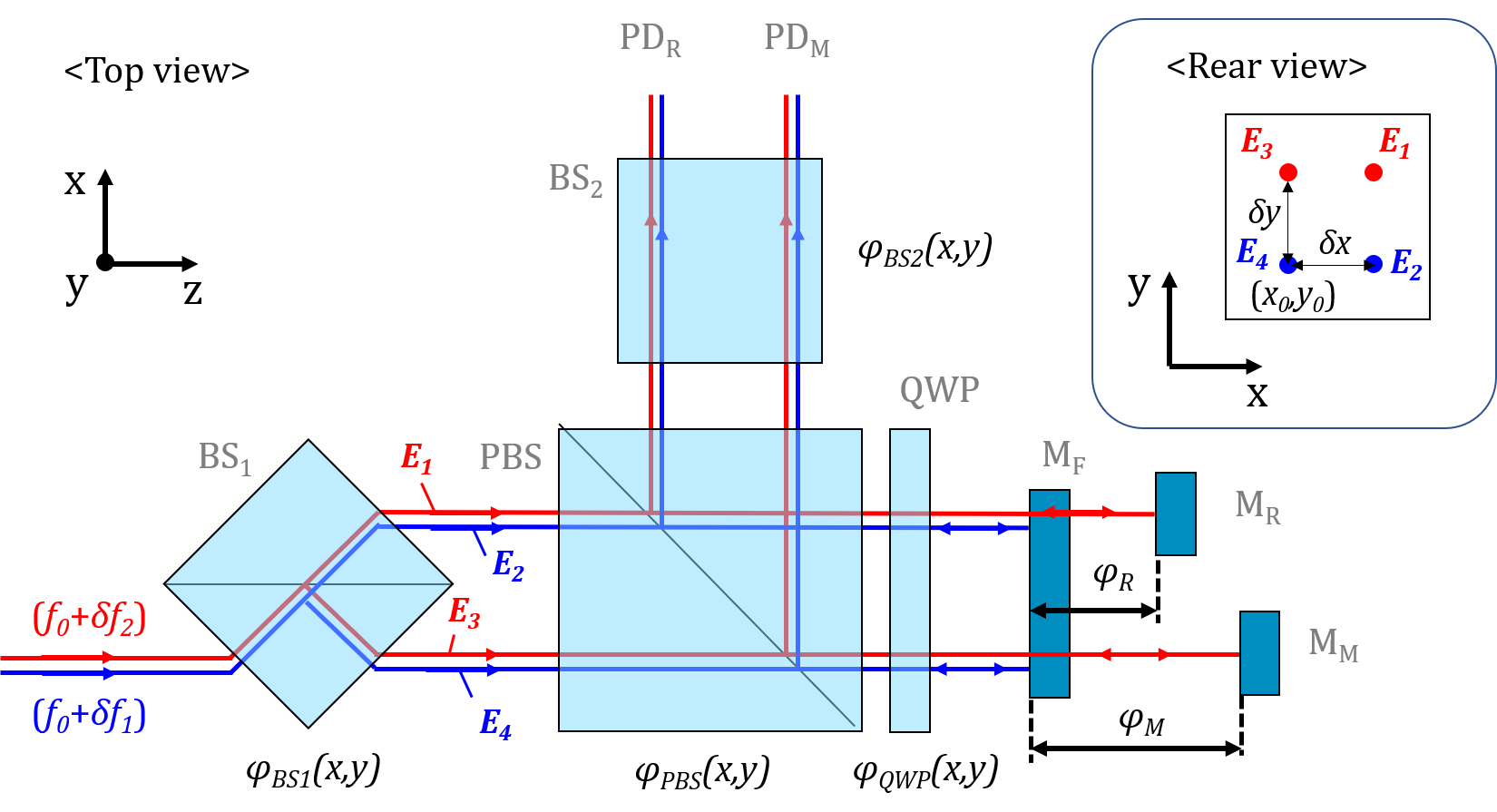}}
\caption{Top view of the proposed differential heterodyne laser interferometer with spatial beam separation shown in \Cref{Fig1}. The inset indicates the rear view of the interferometer with x-y coordinates.}
\label{Fig2}
\end{figure}

\Cref{Fig2} shows the top and rear view of the interferometric setup shown in \Cref{Fig1}. Since the two beams are delivered by two separated fiber collimators connected to polarization-maintaining fibers (PMFs), each beam has different starting phase terms, $\phi_1$ and $\phi_2$ associated to them. The two starting beams are vertically separated and are horizontally split by $\textnormal{BS}_1$, creating the four resulting beams that traverse the entire interferometer. The beams can be expressed as
\begin{align}
\label{eq1}
\begin{split}
E_1 &= E_0 \exp{[i2\pi (f_0 +\delta f_2)t+\phi_2+\phi_t(x_0+\delta x, y_0+\delta y)+\phi_R},\\
E_2 &= E_0 \exp{[i2\pi (f_0 +\delta f_1)t+\phi_1+\phi_t(x_0+\delta x, y_0)},\\
E_3 &= E_0 \exp{[i2\pi (f_0 +\delta f_2)t+\phi_2+\phi_t(x_0, y_0+\delta y)+\phi_M},\\
E_4 &= E_0 \exp{[i2\pi (f_0 +\delta f_1)t+\phi_1+\phi_t(x_0, y_0)},
\end{split}     
\end{align}

\noindent
where $E_0$ is the common electrical field amplitude of the beams, assuming it is the same for all four beams due to their common source. The term, $\phi _t$, is the total phase noise caused by contributions from $\textnormal{BS}_1$ ($\phi _{BS1}$), PBS ($\phi _{PBS}$), QWP ($\phi _{QWP}$) and $\textnormal{BS}_2$ ($\phi _{BS2}$) including optical paths in air and as a function of lateral x-y coordinates, while $\phi _R$ and $\phi _M$ denote the phases by the reference and measurement paths, respectively. The terms ($x_0$,$y_0$) are the lateral coordinates of $E_4$,  and the horizontal and vertical separations are denoted as $\delta _x$ and $\delta _y$ as illustrated in the inset of \Cref{Fig2}. 

In this case, we assume that temperature fluctuations are not common-mode, which means that $\phi _t$ can be divided into two phase noise contributions, $\phi _{tx}$ and $\phi _{ty}$ corresponding to each axis. Then, the interference signals of the reference ($I_R$) and the measurement ($I_M$) can be calculated as 
\begin{align}
\label{eq2}
\begin{split}
I_R &= I_0(1+\cos(2\pi \delta ft+\phi _R+\Delta \phi _{ty})),\\
I_M &= I_0(1+\cos(2\pi \delta ft+\phi _M+\Delta \phi _{ty})),\\
\end{split}     
\end{align}

\noindent
where $I_0$ is the nominal intensity and $\delta f$ is the heterodyne beat frequency, defined as ($\delta f_1-\delta f_2$). The term $\Delta \phi _{ty}$ is the phase noise difference along the y-direction, denoted as $[\phi_{ty}(y_0+\delta y)-\phi_{ty}(y_0)]$ while the phase noise ($\Delta \phi _{tx}$) along the x-direction is absent from \Cref{eq2} because $\phi _{tx}(x_0+\delta x,y_0)$ is the common phase term between $E_1$ and $E_2$, leading to the expression of $I_R$. The term $\phi _{tx}(x_0,y_0)$ is common in $E_3$ and $E_4$ for $I_M$, which is obtained from \Cref{eq1}. Moreover, $\Delta \phi _{ty}$ is also the common phase term in $I_R$ and $I_M$, and cancels in the differential phase measurements between $I_R$ and $I_M$. Under the assumptions leading to the results expressed in \Cref{eq2}, temperature fluctuations will be significantly attenuated in the interferometer while maintaining a simple layout and few optical components. Such a simple and compact laser interferometer can be readily utilized for displacement measurements in a wide variety of research and industrial applications. 

\subsection{Elimination of Periodic Errors}
Periodic errors in heterodyne interferometry are mainly caused by frequency or polarization mixing amongst the interfering beams. To prevent this, we used two AOFSs (not shown in \Cref{Fig1}), on their first diffraction order, which are driven by slightly different RF frequencies ($\delta f_1$ and $\delta f_2$) and fiber-coupled. Moreover, by using PMFs and individual fiber collimators, these two beams have the same polarization and are launched into the interferometer with a physical separation between their optical axes. Periodic errors can thus only come to place by surface reflections at optical components, which can be mitigated by an intentional and well-controlled slight misalignment.

\section{Experimental Results}
\subsection{System Construction}

To verify the performance of our interferometer, we used a commercial frequency stabilized HeNe laser (HRS015B@thorlabs) with $\pm 1$~MHz frequency stability over a time period of 1 minute. The laser output was divided into two beams, each passing through an AOFS (1205C-2@Isomet) operating at 80 and 85 MHz respectively. Each positive first order of diffraction was coupled into PMF and delivered to the optical interferometer assembly built by commercial polarizing and non-polarizing beam splitters. After aligning the optical components and mirrors, we obtained two beat notes, each with a contrast above 80\%. We used a commercial phasemeter (Moku:Lab@Liquid Instrument) to measure the heterodyne signals emerging from the photodetectors, $\textnormal{PD}_\textnormal{R}$ and $\textnormal{PD}_\textnormal{M}$. We installed the interferometer on an optical table and operated it in air. We enclosed the optical setup in an insulating box consisting of sponge foam sheets to reduce the impact of environmental noises.  The Moku:Lab phasemeter offers discrete selectable sampling frequency options and thus the sampling frequencies chosen during experimentation are chosen based on the desired resolution and duration of each measurement.

\subsection{Interferometer Stability Tests}

To measure the stability and self-noise of the interferometer, we used a single flat mirror instead of individual, mechanically decoupled reflective surfaces for $\textnormal{M}_\textnormal{F}$, $\textnormal{M}_\textnormal{R}$ and $\textnormal{M}_\textnormal{M}$.

\subsubsection{Short-term stability}
\begin{figure}[htbp]
    \centering
    \begin{subfigure}[b]{\linewidth}
    \centering
    \fbox{\includegraphics[width=.95\linewidth]{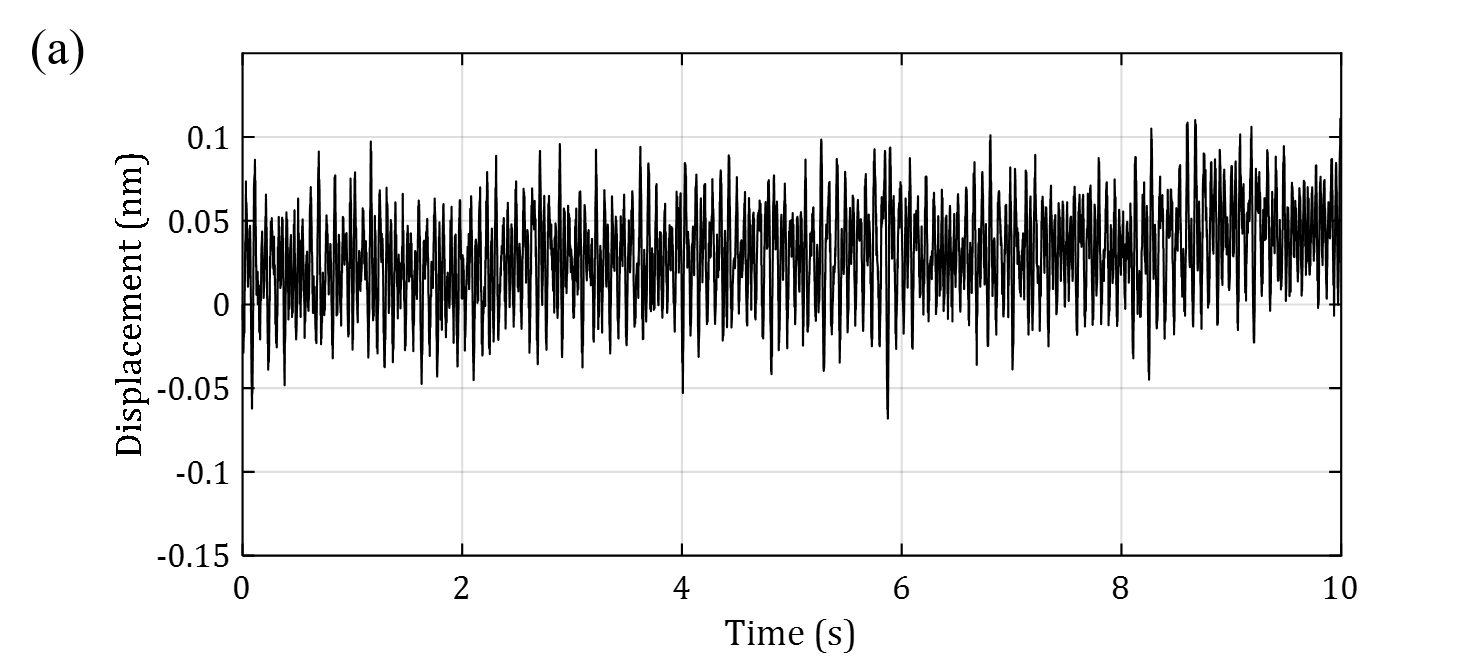}}\phantomsubcaption{}\label{Fig3a}
    \end{subfigure}
    \begin{subfigure}[b]{\linewidth}
    \centering
    \fbox{\includegraphics[width=.95\linewidth]{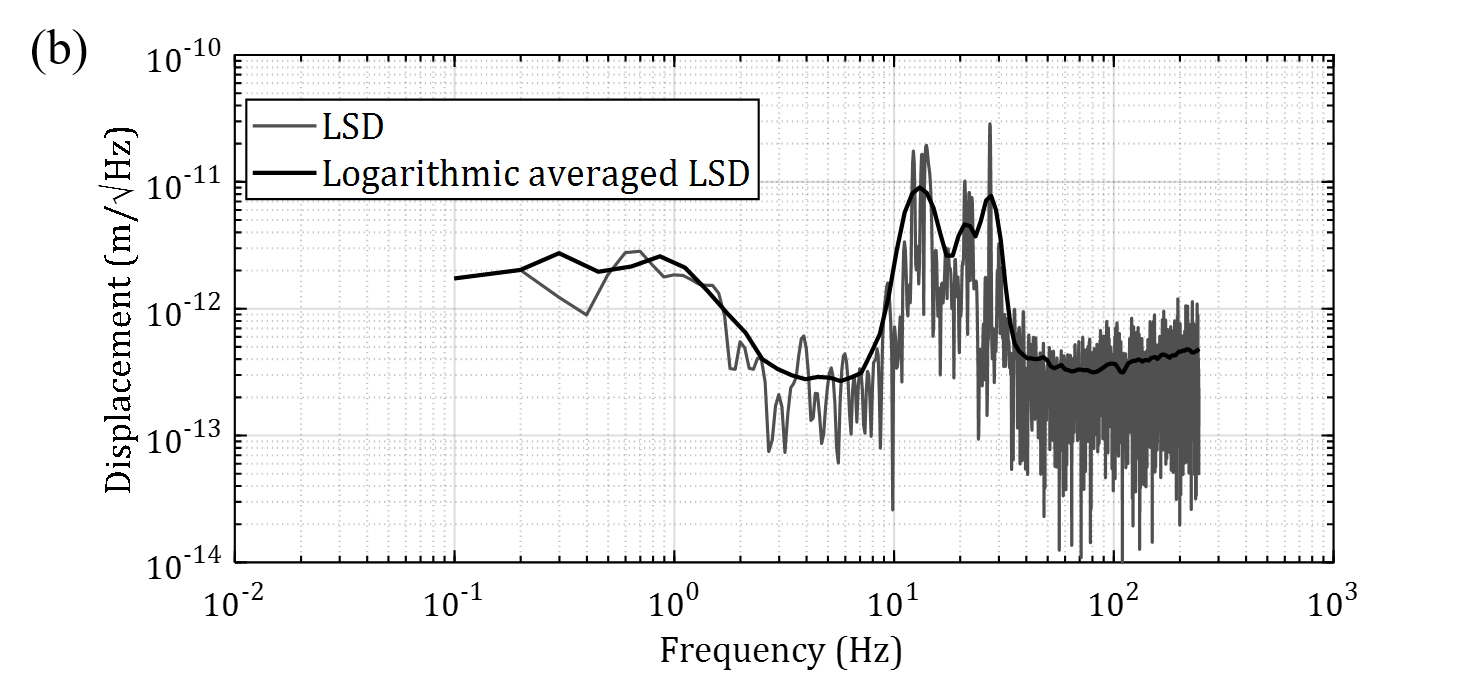}}\phantomsubcaption{}\label{Fig3b}
    \end{subfigure}
    \caption{(a) Displacement measurement results during 10 s for the short-term stability test and (b) linear spectral density of (a).}
    \label{Fig3}
\end{figure}

We measured the short-term stability over a period of 10~s with a stationary single mirror. We set the phasemeter to a phase sampling frequency of 488~Hz to investigate the mechanical stability of the interferometer below 500 Hz. \Cref{Fig3a,,Fig3b} show the time-series of the measured displacement and its fluctuations as a linear spectral density (LSD), respectively. We computed the displacement standard deviation to be 28~pm. Evidently, the main noise sources of the system seem to be mechanical instabilities (peaks) in the frequency range from 10~Hz to 30~Hz, which we attributed to acoustic noise in the laboratory. 

\subsubsection{Long-term stability}
\begin{figure}[htbp]
    \centering
    \begin{subfigure}[b]{\linewidth}
    \centering
    \fbox{\includegraphics[width=\linewidth]{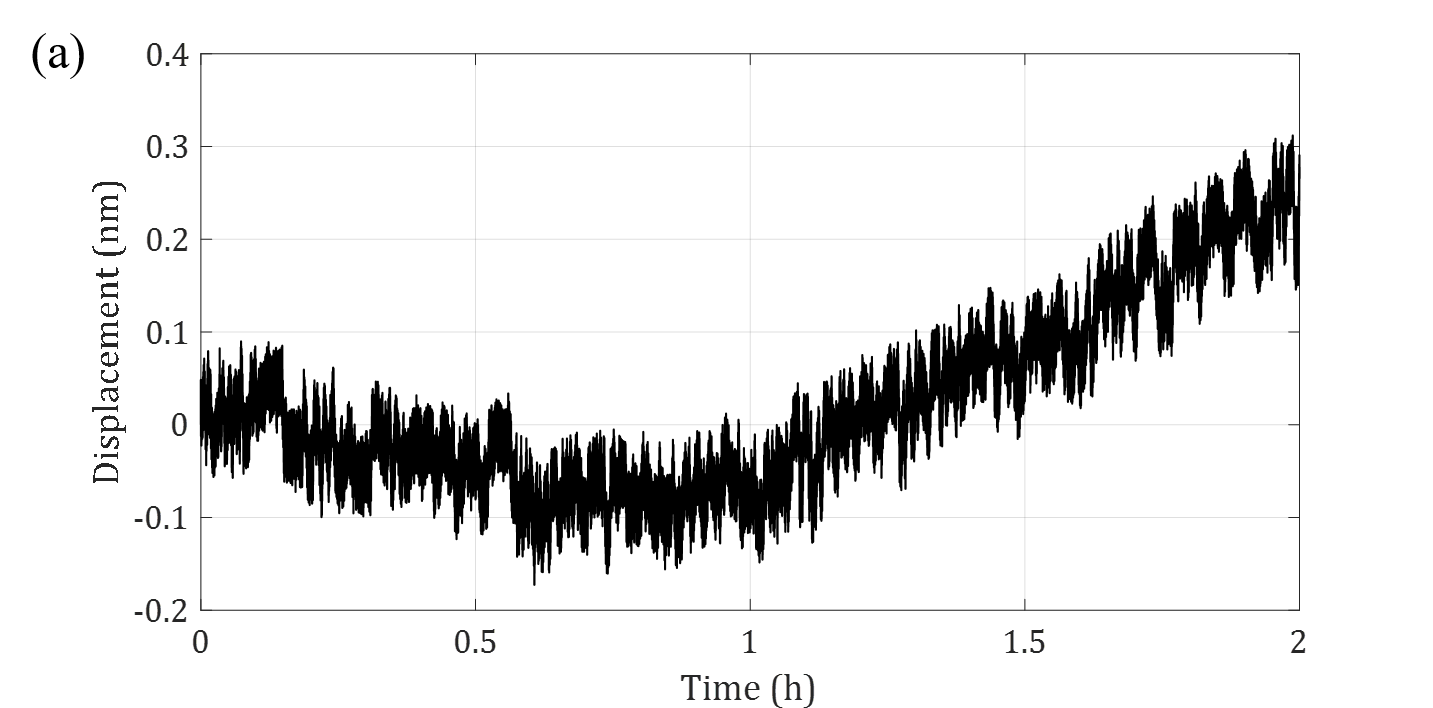}}\phantomsubcaption{}\label{Fig4a}
    \end{subfigure}
    
    \begin{subfigure}[b]{\linewidth}
    \centering
    \fbox{\includegraphics[width=\linewidth]{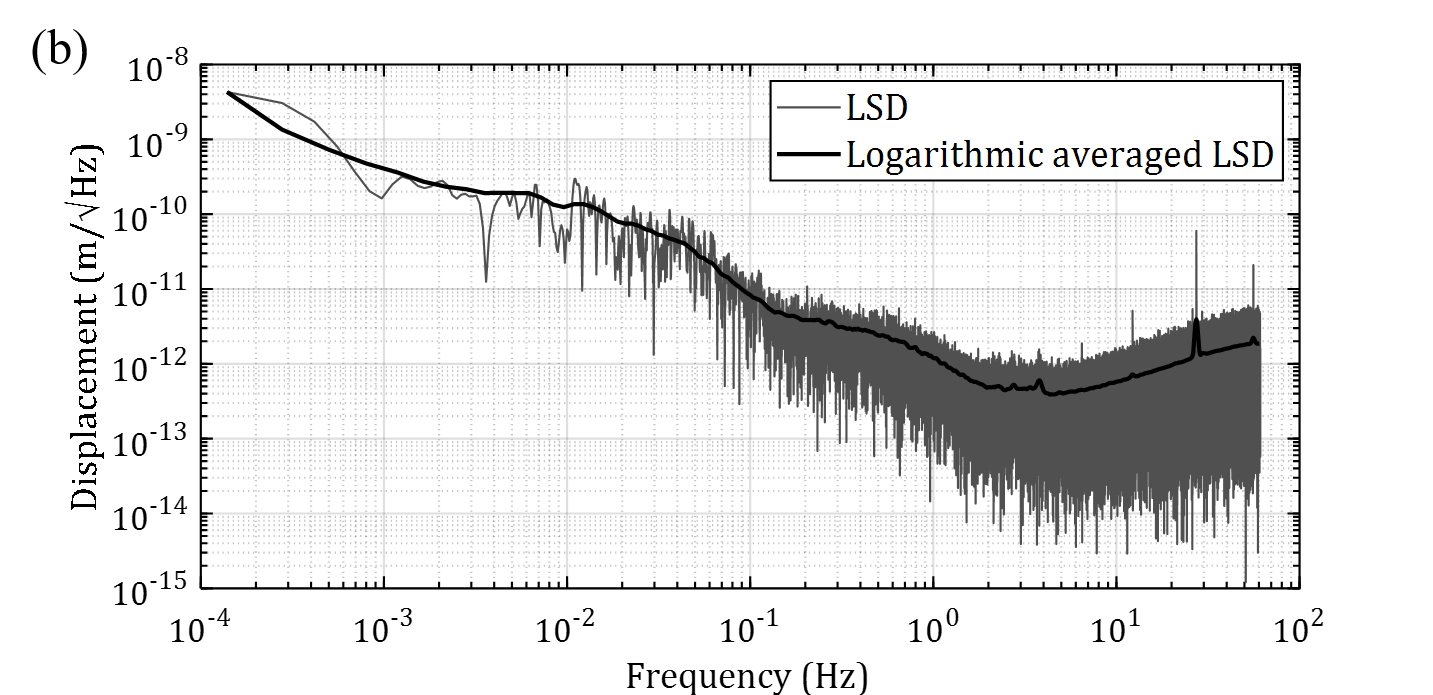}}\phantomsubcaption{}\label{Fig4b}
    \end{subfigure}
    \caption{(a) Displacement measurement results during 2 h and (b) linear spectral density (LSD) of (a).}
    \label{Fig4}
\end{figure}

We conducted displacement measurements over a period of 2~hours at a phase sampling frequency of 122~Hz, as shown in  \Cref{Fig4a}, to estimate the long-term stability. A lower sampling frequency was chosen to observe the long-term trend while avoiding large data lengths. The measurement results show a long-term drift of approximately 0.3~nm. Our temperature measurements show that the origin of this drift is residual temperature fluctuations coupling into the interferometric phase measurement. \Cref{Fig4b} shows the linear spectral density (LSD) of these measurements. 

\begin{figure}[htbp]
    \centering
    \begin{subfigure}[b]{\linewidth}
    \centering
    \fbox{\includegraphics[width=.95\linewidth]{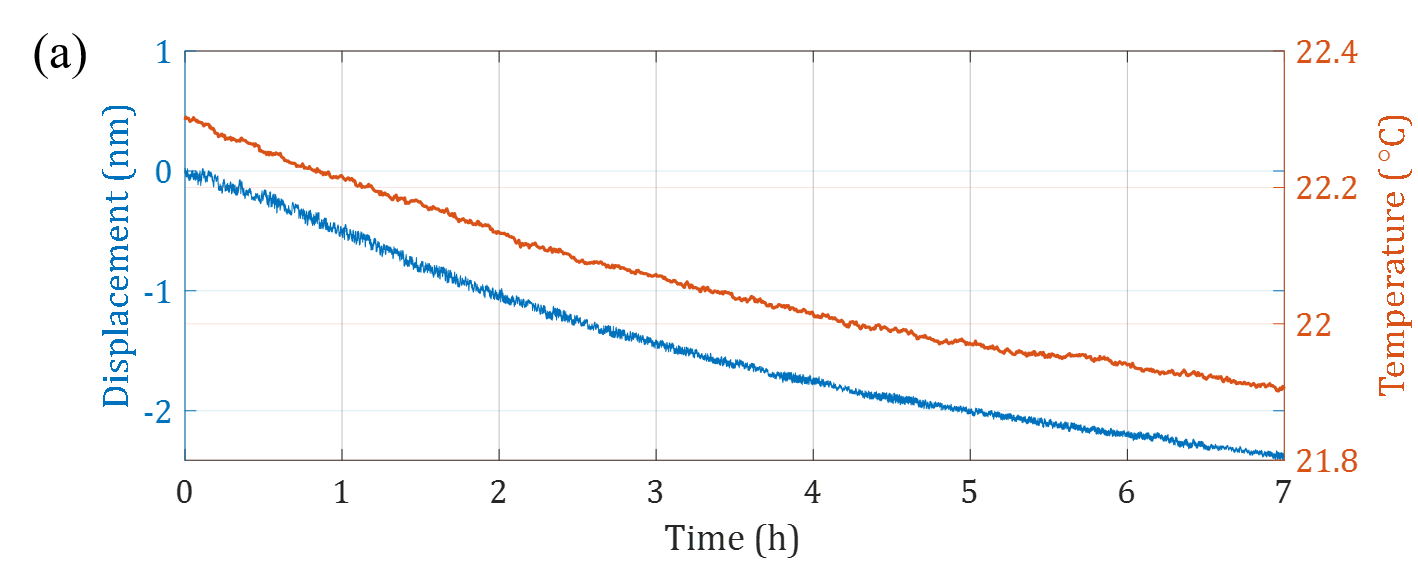}}\phantomsubcaption{}\label{Fig5a}
    \end{subfigure}

    \begin{subfigure}[b]{\linewidth}
    \centering
    \fbox{\includegraphics[width=.95\linewidth]{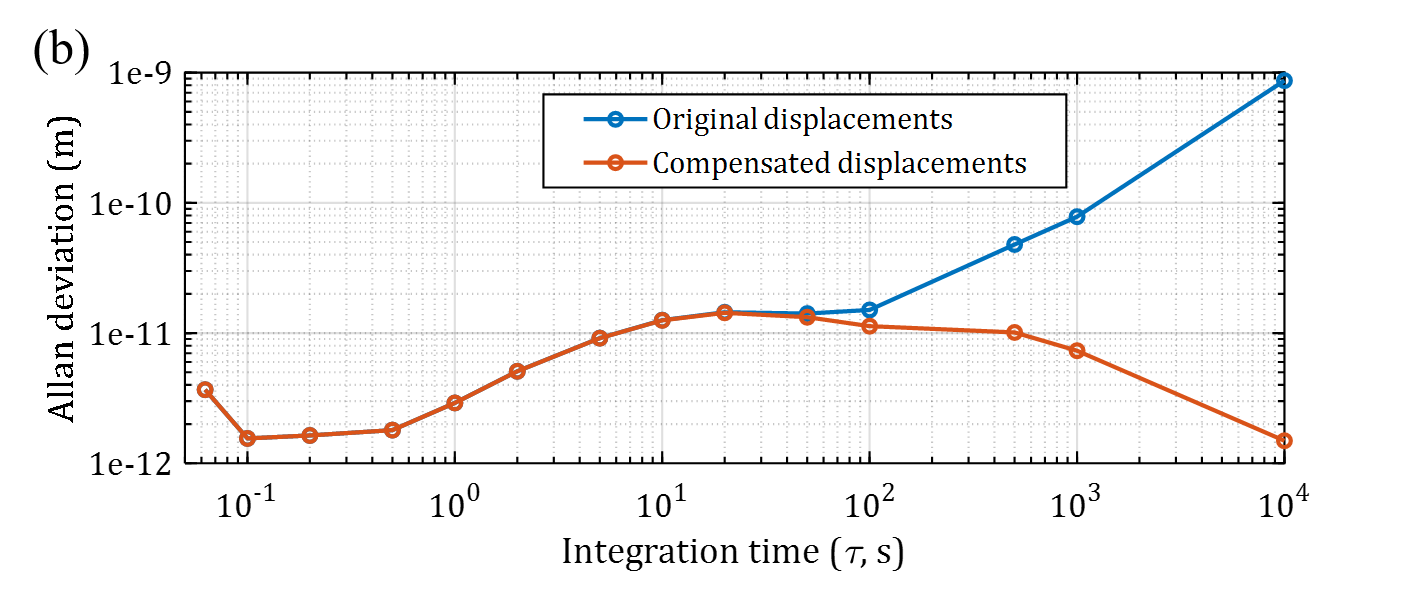}}\phantomsubcaption{}\label{Fig5b}
    \end{subfigure}
    \caption{(a) Displacement measurement results and temperature variation during 7~h (b) Allan deviations of original and compensated displacements for the long-term stability.}
    \label{Fig5}
\end{figure}

As expected, temperature fluctuations become dominant in the low frequency regime, while acoustic noise appears in the range from 10~Hz to 60~Hz, similarly to the short-term stability result in \Cref{Fig3}. To verify the long-term thermal drift of the measurement results, we used a home-built temperature sensor â€“ consisting of an NTC (103) thermistor and a 10~k$\Omega$ resistor â€“ to monitor the ambient temperature over a period of 7 hours and sampled at a frequency of 30.5~Hz, whiles simultaneously measuring the interferometer phase. \Cref{Fig5a} presents the temperature variation and the resulting displacement measurement. The long-term drift of the displacement is clearly driven by temperature variations, yielding a drift of approximately 2.0~nm over a temperature range between 22.3{\degree}C to 21.9{\degree}C. The temperature coupling factor, in this case, was  5.0~nm/K. As a comparison, a high stability commercial differential plane mirror interferometer has approximately 10~nm/K as quoted by the manufacturer~\cite{23Zygo}. 

To evaluate the long-term stability, we computed the Allan deviation for this displacement measurement run as illustrated in \Cref{Fig5b}. This shows a system stability of 3~pm over an integration time of 1~s. As a result of the temperature fluctuations, the Allan deviation clearly increases over longer integration times as shown by the blue trace curve. In addition, having determined the temperature coupling coefficient into our interferometric readout, it is possible to compensate our displacement measurements by subtracting the measured temperature fluctuations. This result is shown in the orange trace, labelled 'compensated'. In this case, the Allan deviation shows stability levels of 3~pm over 1~s, same as the original uncompensated data, 10~pm over 100~s, and slightly below 2~pm at integration times of 10,000~s.

The temperature coupling mechanism in our setup is attributed to the pointing instability between the two beams because the fiber collimators launching the two input beams into the interferometer were aligned with separate kinematic mounts. Therefore, temperature variations can cause misalignments of the two parallel beams, which subsequently leads to displacement drifts.

\subsection{Displacement Measurements}

To measure real physical displacements with our interferometer, we aligned three individual and mechanically decoupled mirrors separately while we attached $\textnormal{M}_\textnormal{M}$, the target, to a piezoelectric transducer (PZT). The nominal distance between $\textnormal{M}_\textnormal{R}$ and $\textnormal{M}_\textnormal{M}$ was approximately 20~mm. We conducted displacement measurements over a time period of 2~minutes with a phasemeter sampling frequency of 488~Hz.

\subsubsection{Frequency response at a fixed position}
\begin{figure}[htbp]
\centering
\fbox{\includegraphics[width=.95\linewidth]{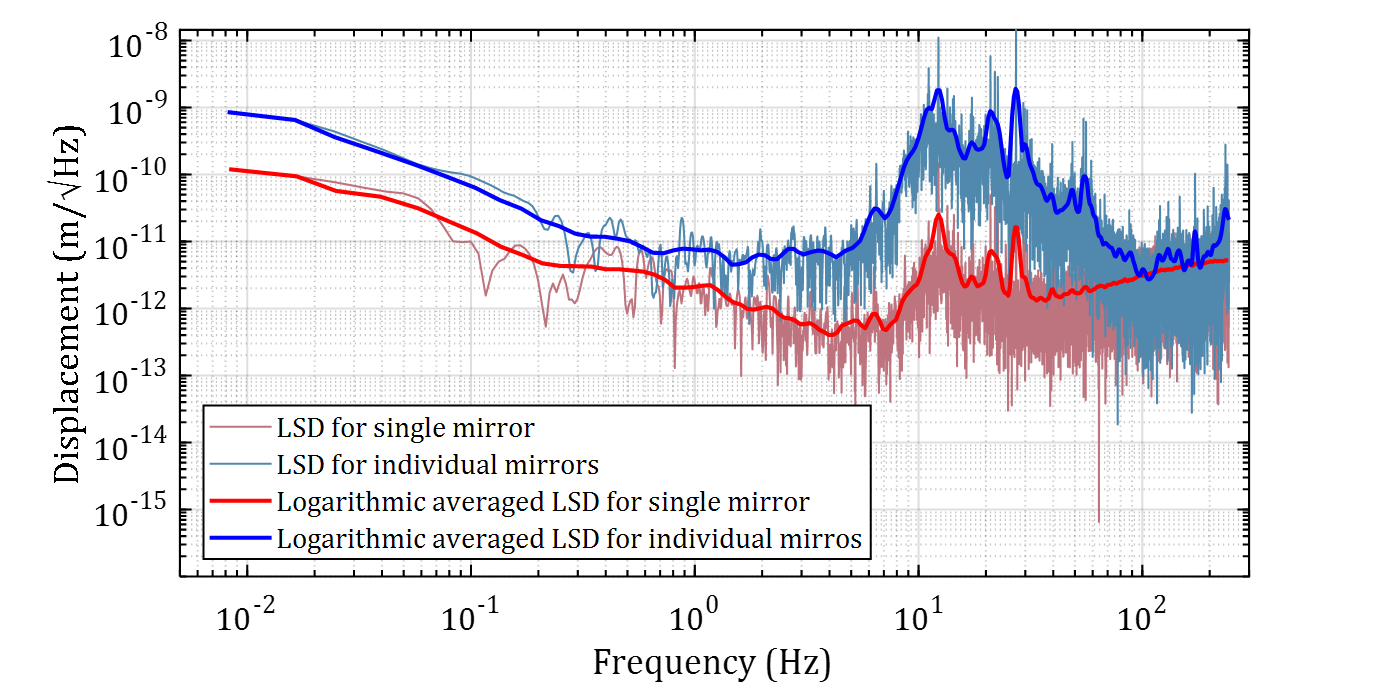}}
\caption{Linear spectral density (LSD) of fixed individual mirror motions compared to that of the single mirror motion.}
\label{Fig6}
\end{figure}

Before measuring the motion of $\textnormal{M}_\textnormal{M}$ when driven by a PZT, we first conducted a baseline measurement with $\textnormal{M}_\textnormal{M}$ stationary, which is shown in \Cref{Fig6}. For comparison purposes, we also show the data measured with a single mirror. Compared to  \Cref{Fig3b}, the displacement noise floor increases over the entire frequency range, especially between 10~Hz and 100~Hz. This can be attributed to vibrations and acoustic noises, that are coupled more strongly with individual mirrors, $\textnormal{M}_\textnormal{R}$ and $\textnormal{M}_\textnormal{M}$, as opposed to using a single common reflecting surface.

\begin{figure}[htbp]
    \centering
    \begin{subfigure}[b]{\linewidth}
    \centering
    \fbox{\includegraphics[width=.95\linewidth]{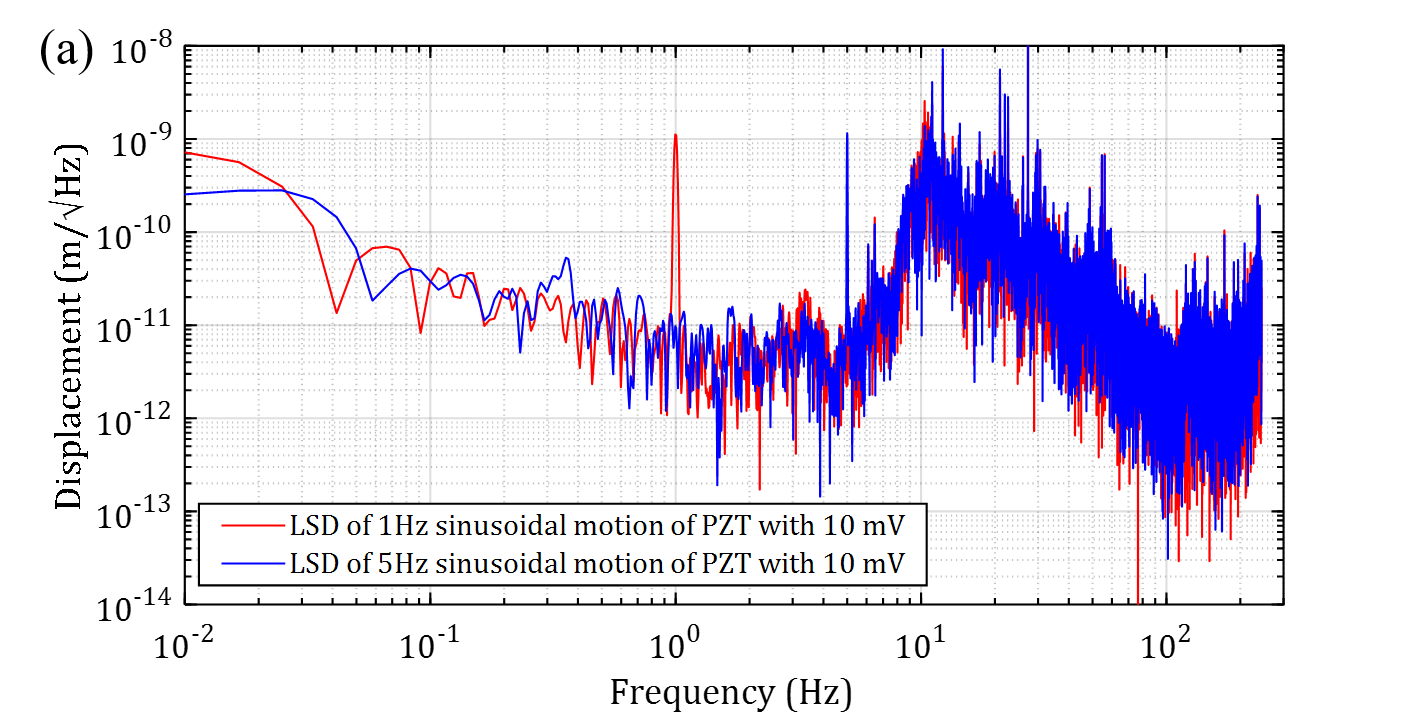}}\phantomsubcaption{}\label{Fig7a}
    \end{subfigure}

    \begin{subfigure}[b]{\linewidth}
    \centering
    \fbox{\includegraphics[width=.95\linewidth]{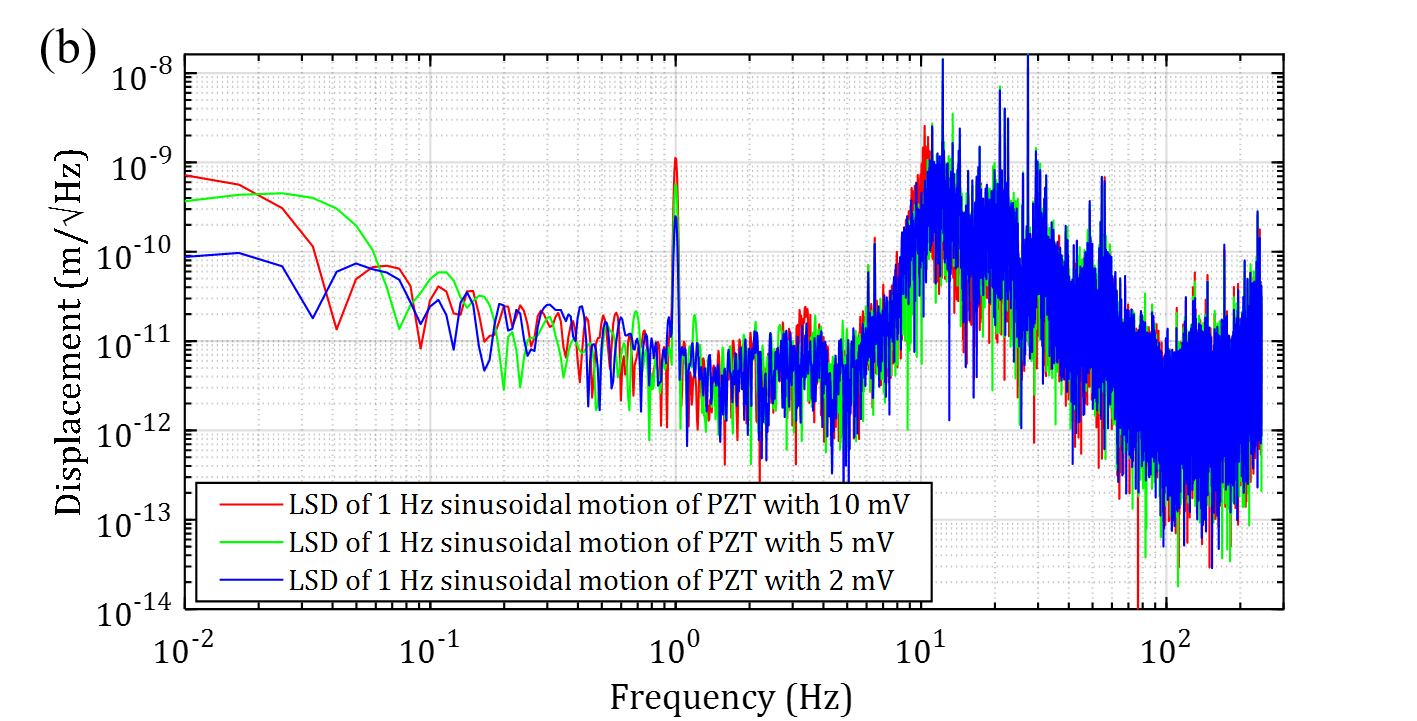}}\phantomsubcaption{}\label{Fig7b}
    \end{subfigure}
    \caption{(a) Linear spectral density (LSD) plots of 1~Hz and 5~Hz PZT motions with 10~mV and (b) Linear spectral density (LSD) plots of 1~Hz PZT motions with 10~mV, 5~mV and 2~mV.}
    \label{Fig7}
\end{figure}

The PZT was driven by sinusoidal signals at 1~Hz and 5~Hz with a peak to peak amplitude ($\textnormal{V}_\textnormal{pp}$) of 10~mV. We computed the LSDs of the interferometer data containing these excitations, which are shown in  \Cref{Fig7a}. The peaks at 1~Hz and 5~Hz are clearly visible and stand out over the measurement noise floor, which is consistent with the fixed mirror $\textnormal{M}_\textnormal{M}$ measurement shown in \Cref{Fig6}.  In addition, we varied the amplitude of the excitations injected through the PZT to verify the response of the interferometer. We adjusted the amplitudes to 10~mV, 5~mV and 2~mV at 1~Hz, and their response is clearly visible in the LSD plot shown in \Cref{Fig7b}, with displacement peak values of 0.57~nm, 0.30~nm and 0.13~nm, respectively. 

\subsection{Comparison Displacement Measurements to a Commercial DMI System}
For these displacement measurements, we attached $\textnormal{M}_\textnormal{M}$ to a commercial PZT stage (MAX312D@Thorlabs) and set our phasemeter to a sampling frequency of 30.5~Hz, which is similar to the sampling frequency settings in the commercial DMI. We used a commercial DMI system (XL80@Renishaw) and attached its measurement retroreflector to the same PZT stage to measure its dynamics and compare the results obtained with our interferometer. The measurement results of the commercial DMI were averaged over 0.1~ms and the sampling frequency was 20~Hz, which is close to the sampling frequency of our phasemeter. 

\subsubsection{Displacement measurement comparison}

\begin{figure}[htbp]
    \centering
    \begin{subfigure}[b]{\linewidth}
    \centering
    \fbox{\includegraphics[width=.95\linewidth]{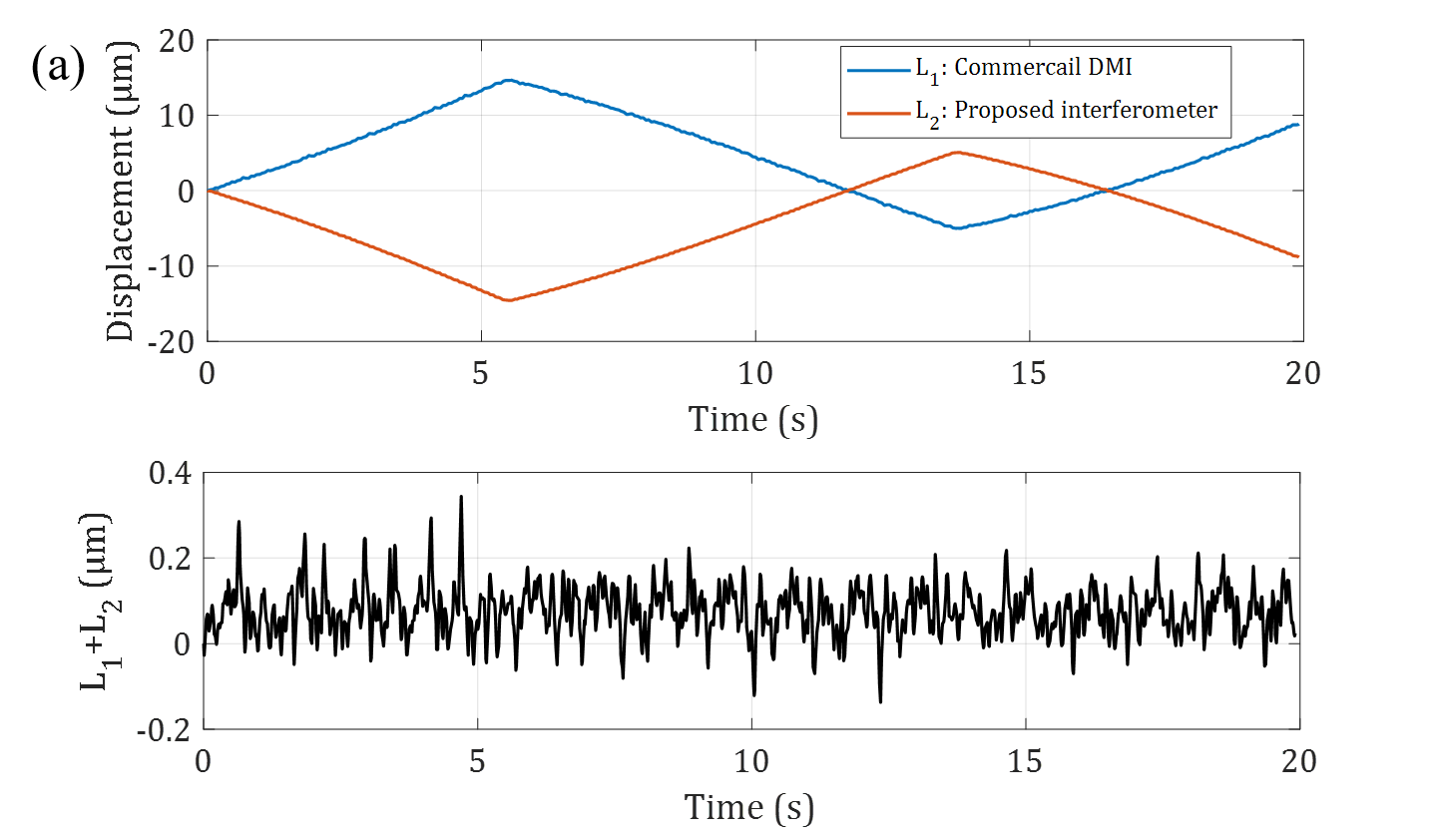}}\phantomsubcaption{}\label{Fig8a}
    \end{subfigure}

    \begin{subfigure}[b]{\linewidth}
    \centering
    \fbox{\includegraphics[width=.95\linewidth]{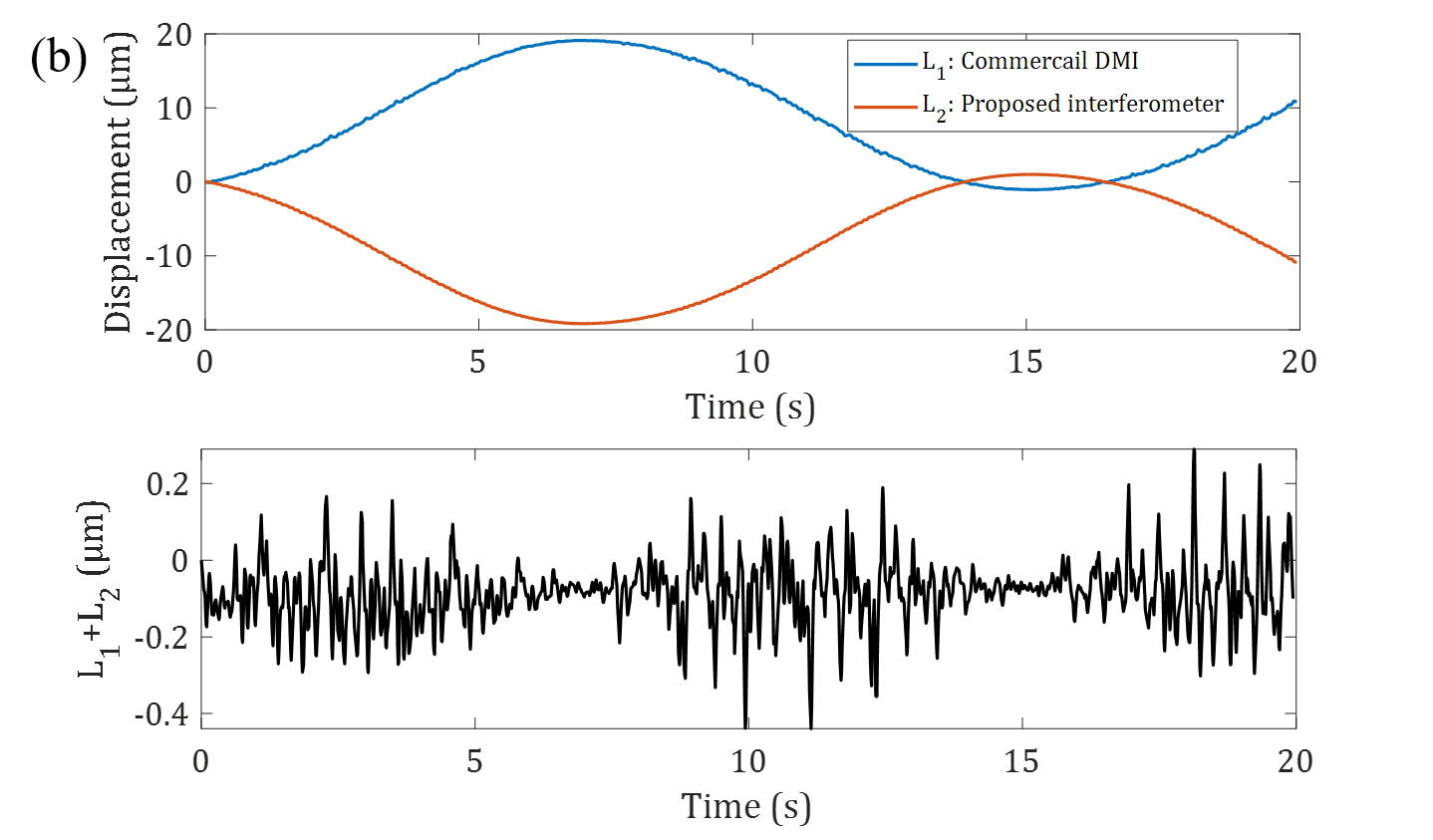}}\phantomsubcaption{}\label{Fig8b}
    \end{subfigure}
    \caption{Displacement measurement results of the commercial DMI ($\textnormal{L}_1$), the proposed interferometer ($\textnormal{L}_1$) and ($\textnormal{L}_1+\textnormal{L}_2$) for (a) a triangle and (b) a sinusoidal motion with 0.06~Hz.}
    \label{Fig8}
\end{figure}

Figure 8 shows the displacement measurement results obtained with our interferometer and the commercial system when the PZT stage was moved with triangular and sinusoidal waves at a frequency of 0.06~Hz and an amplitude of 75~$\textnormal{V}_\textnormal{pp}$, which corresponds to a PZT stage excursion of approximately 20~{\textmu}m over 20~s. As shown in \Cref{Fig8a,,Fig8b}, the two systems measure similar results, but with opposite signs, as is expected from measuring targets on the stage mounted to reflect counterpropagating beams. The standard deviations of the difference between these measurements amount to 55~nm in the triangular case, and 81~nm in the sinusoidal. The main reasons for the differences between the two results were the distinct step responses corresponding to the stage motions. The stage used in this investigation was operated in open-loop with 20~nm resolution steps. The measurement results differ from each other at the transitions of the step-wise motion. \Cref{Fig8b} shows this effect where several spikes appear during the motion while the differences were reduced at the maxima and minima regions. Furthermore, it is important to mention that contributions to these errors may come from each interferometer measuring a different target mounted on a common PZT stage, since the commercial DMI requires a dedicated retroreflector, and not a normal-incidence mirror as our prototype interferometer.

\begin{figure}[htbp]
\centering
\fbox{\includegraphics[width=.95\linewidth]{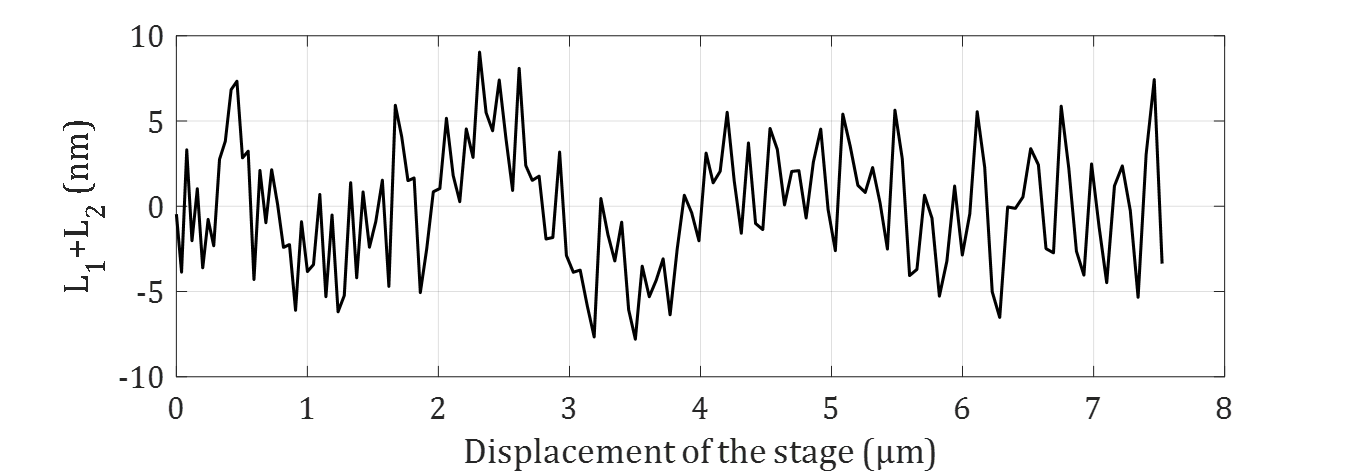}}
\caption{($\textnormal{L}_1+\textnormal{L}_2$) for 20 s time-averaged displacements. }
\label{Fig9}
\end{figure}

It is clear that the two systems show better agreement in their results, when measuring at fixed stationary positions as illustrated in \Cref{Fig9}. In this case, the standard deviation between the systems is 3.6~nm, which can be attributed to the higher noise floors of the commercial system.

\subsubsection{Periodic error detection}
\begin{figure}[htbp]
    \centering
    \begin{subfigure}[b]{\linewidth}
    \centering
    \fbox{\includegraphics[width=.95\linewidth]{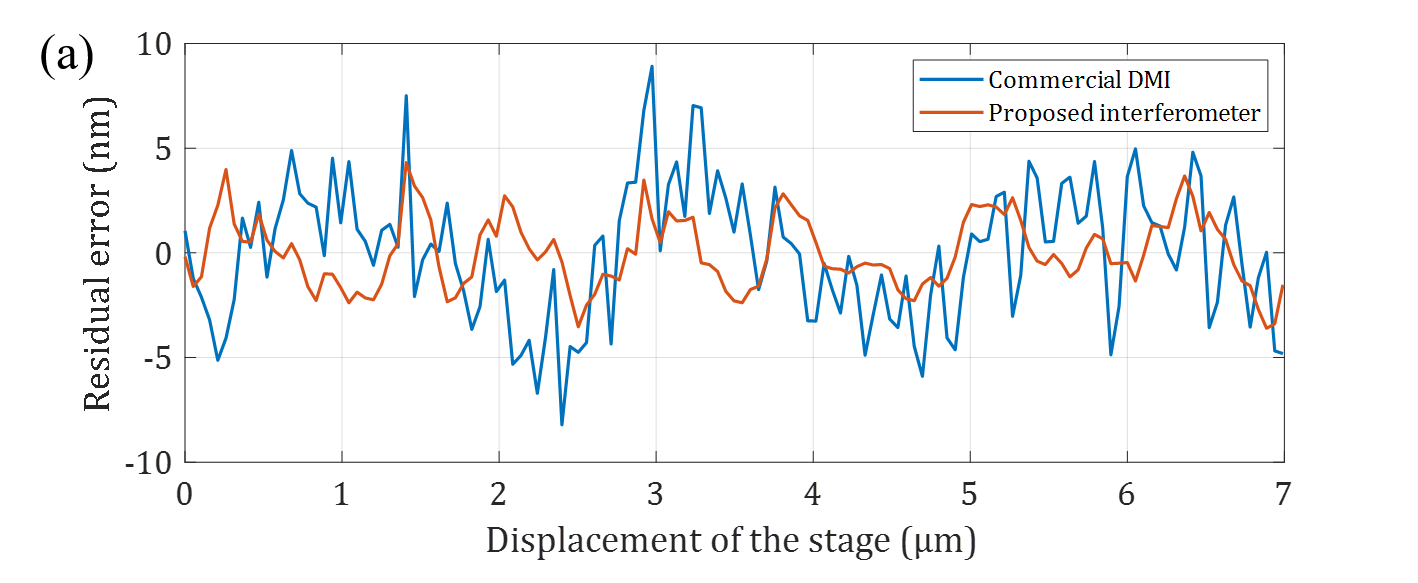}}\phantomsubcaption{}\label{Fig10a}
    \end{subfigure}

    \begin{subfigure}[b]{\linewidth}
    \centering
    \fbox{\includegraphics[width=.95\linewidth]{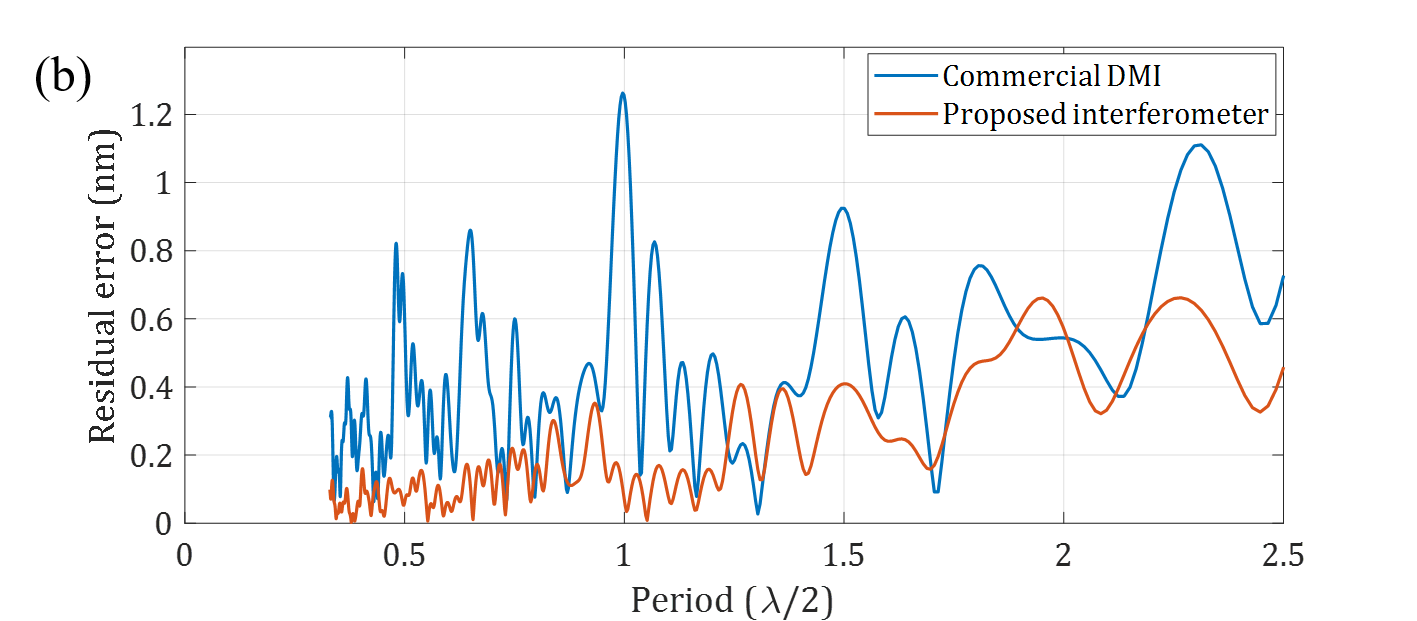}}\phantomsubcaption{}\label{Fig10b}
    \end{subfigure}
    \caption{(a) Displacement measurement results and (b) the periodicity of the residual errors for the commercial DMI and the proposed interferometer.}
    \label{Fig10}
\end{figure}

We evaluated the periodic errors of the interferometers over the residual position errors after applying a polynomial curve fit to the stage motion. To reduce the transient errors of the stage translation steps, we measured displacements at steps every 50~nm during 5~s and averaged over a 7~{\textmu}m measurement range. \Cref{Fig10} shows the measurement results for the nonlinear errors. We fitted the linear motion of the stage and the measurement results with a six order polynomial~\cite{24Lawall}, and obtained the residuals by assuming a continuous translation of the stage that would allow us to extract the nonlinear errors, shown in \Cref{Fig10a}. 

We computed the Fourier transform of these measurements to visualize better the content of periodic nonlinearities in the interferometers. \Cref{Fig10b} presents these spectra, plotted as fractional half-wavelengths ($\lambda /2$). The commercial DMI system has a first order periodic error of 1.25~nm, despite of the fact that it is a homodyne laser interferometer. This, however, is likely caused by the quadrature detection scheme that uses polarizing optics to measure the interferometer phase and its sign, which leads to polarization mixing~\cite{25Eom}. Conversely, the data from our prototype heterodyne interferometer shows no detectable first, second, or higher order periodic errors.

\section{Discussion}
As shown in \Cref{Fig10}, we demonstrated that our prototype heterodyne interferometer has no detectable periodic errors caused by frequency or polarization mixing to the displacement measurement resolutions presented. 

\begin{figure}[htbp]
\centering
\fbox{\includegraphics[width=.95\linewidth]{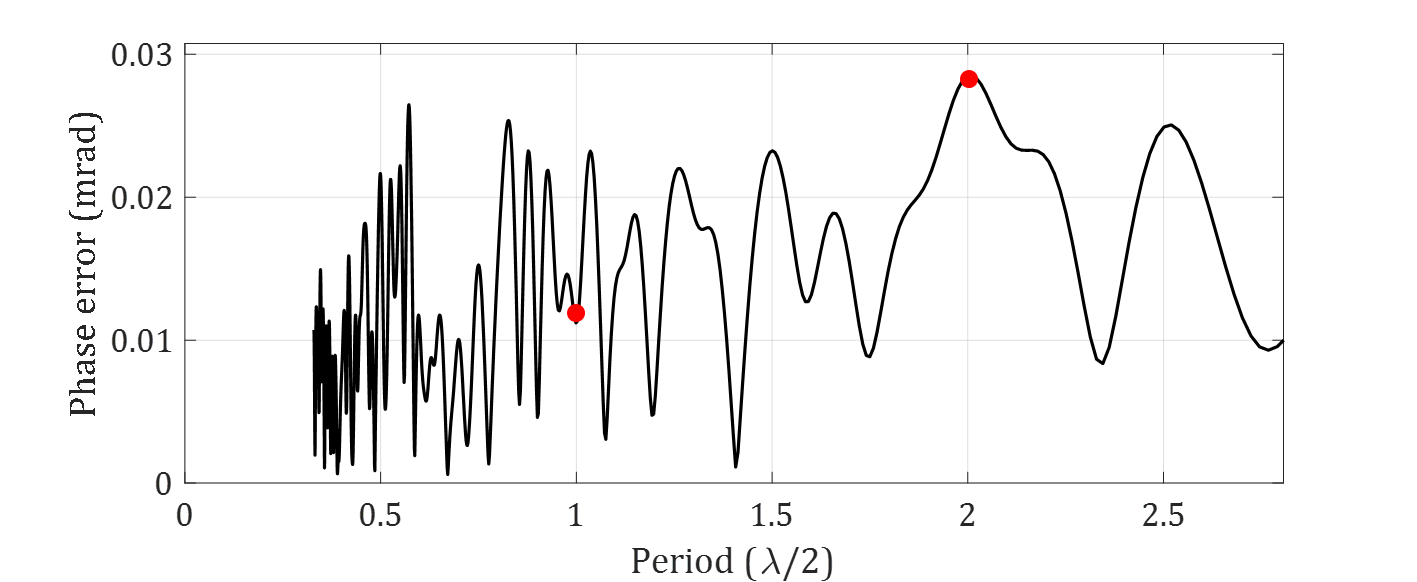}}
\caption{Phase error calculated by the amplitude method.}
\label{Fig11}
\end{figure}

To further investigate the presence of periodic errors in our interferometer, we analyzed the relative amplitude changes ($\Delta R/R$) \cite{19Wu} of the heterodyne signals and calculated their effect on the measured phase errors, as seen in \Cref{Fig11}. Here, the phase error corresponding to the first order periodic errors is negligible since its level is around the overall baseline of the trace, while that of the second order reaches a level of 0.028~mrad. By using this analysis method for the relative heterodyne amplitudes, we can estimate periodic error levels at 3.5~pm and 9~pm for the first and second orders, respectively. 

Another factor to consider is the long-term power stability of the optical source and the AOFS output beams. To determine the impact of this, we conducted a simulation analysis that showed that intensity fluctuations at levels of 1\% between two signals caused displacement errors (standard deviation) of approximately 28~pm. Our HeNe laser provides a stable frequency, which reduces intensity fluctuations at the source, but not at the AOFS, which are known to be sensitive to temperature variations. During our long-term experiments shown in \Cref{Fig5}, we observed intensity fluctuations at levels below 2\%. We can achieve further improvements through active intensity stabilization at both the source and the AOFS. 

Lastly, as previously mentioned, we conducted all of our measurements presented here in air, under environmental conditions that were not being controlled. Once we install our interferometer in a vacuum chamber with adequate thermal shielding (around 1 mK/$\sqrt{\textnormal{Hz}}$) and under well-controlled environmental conditions, we anticipate a significant performance enhancement since most thermal and acoustic disturbances, and the resulting misalignment errors, will be significantly reduced.

\section{Summary}

We  presented and experimentally demonstrated a compact differential heterodyne laser interferometer without periodic errors. We designed this interferometer with an intentional spatial separation between the optical axes of two frequency-shifted beams, thus preventing frequency and polarization mixing. The interferometer pathlengths are designed for a high common-mode rejection ratio to significantly reduce the effect of environmental disturbances, allowing their mitigation through differential measurements between reference and measurement interferometer arms. We conducted functional displacement measurements in comparison with a commercial DMI. These measurements confirmed that nonlinear periodic errors were not detectable in our heterodyne interferometer. Performance experiments of short and long-term stabilities show displacement noise floors of 0.1â€“1~nm/$\sqrt{\textnormal{Hz}}$ at frequencies below 1~mHz and better than 10~pm/$\sqrt{\textnormal{Hz}}$ above 100~mHz. We were able to measure the coupling of room temperature fluctuations to the interferometer displacement measurements and determine an Allan deviation interferometric displacement stability of 3~pm, 10~pm, and 2~pm over integration times of 1~s, 100~s, and 10,000~s, respectively.

\section*{Funding}
This work was supported, in part, by the National Science Foundation (NSF) (PHY-1912106).

\section*{Acknowledgments}
The authors thank Cristina Guzman for reviewing and improving parts of the manuscript.

\section*{Disclosures}
The authors declare that there are no conflicts of interest related to this article.

\bibliography{Reference}
 
\end{document}